\def\ARXIV{1}
\documentclass[11pt]{article}

\ifdefined\ARXIV
  \usepackage[preprint]{acl}
\else
  \usepackage[final]{acl}
\fi

\usepackage{times}
\usepackage{latexsym}
\usepackage[T1]{fontenc}
\usepackage[utf8]{inputenc}
\usepackage{microtype}
\usepackage{inconsolata}
\usepackage{amsmath}
\usepackage{booktabs}
\usepackage{graphicx}
\usepackage{rotating}
\usepackage{fancyvrb}
\usepackage{xspace}

\makeatletter
\@ifpackageloaded{lineno}{%
  \leftlinenumbers%
  \setlength{\linenumbersep}{0.22cm}%
}{}
\makeatother

\newcommand{\method}{E-SENS\xspace}
\definecolor{promptborder}{RGB}{190,95,0}
\definecolor{promptfill}{RGB}{255,248,240}

\title{E-SENS: Exclusion-Sensitive Penalization for\\ Negative-Constraint Retrieval}

\author{Yerang Kim \\
  Independent Researcher \\
  \texttt{hs01151116@korea.ac.kr}
  \And
  Jiyoon Myung \\
  Independent Researcher \\
  \texttt{jiyoon0424@gmail.com}
  \And
  Joohyung Han \\
  Lomin \\
  \texttt{ddan8jh@gmail.com}}

\hypersetup{
  pdftitle={E-SENS: Exclusion-Sensitive Penalization for Negative-Constraint Retrieval},
  pdfauthor={Yerang Kim, Jiyoon Myung, Joohyung Han}
}

\begin{document}
\maketitle

\begin{abstract}
Retrieval-augmented language models can fail to respect negative constraints when the retriever supplies evidence about concepts the user explicitly excluded.
Beyond explicit negation, queries may ask for answers that include one concept while excluding another, or for entities that belong to a category but differ from a closely related instance.
Because the excluded concept still appears in the query text, dense retrievers may assign high similarity to documents about that concept even when the user asks to avoid it.
We introduce \method, a training-free reranking method for negation-sensitive retrieval.
\method extracts a compact trap query for the excluded side and subtracts trap-query similarity from the original-query retrieval score.
On ExcluIR, \method shows a clear recall--violation trade-off across four embedding models and reduces trap retrieval at recall-preserving settings.
\end{abstract}

\section{Introduction}

Users often express information needs by specifying not only what should be retrieved, but also what should be avoided.
A query may ask for documents about one topic while excluding a related entity, or for items that belong to a category but differ from a particular instance.
For example, ``training methods for language models, excluding reinforcement learning'' asks the retriever to preserve the broad topic while avoiding a specific, highly related training paradigm.
In retrieval-augmented generation, returning such excluded material can place constraint-violating evidence directly into the generator context.
This creates an evidence-selection failure: the retriever supplies material that is topically similar to the request but invalid under the user's exclusion constraint.

This setting is difficult for dense retrieval because the excluded concept is both salient and present in the query.
The retriever may assign high similarity to a document precisely because it matches the term the user asked to avoid, reflecting the lexical and semantic signals retained in dense representations \citep{ram-etal-2023-token}.
The core issue is not merely that the query contains negation, but that standard retrievers treat included and excluded terms as undifferentiated semantic evidence.
Negative queries with exclusion constraints therefore require a distinction between inclusion and exclusion semantics that standard retrievers often do not encode reliably \citep{lee2026deotrainingfreedirectembedding}.

Recent benchmarks and systems study this weakness across minimally different negated pairs, realistic exclusion queries, and negative-constraint retrieval \citep{weller-etal-2024-nevir, zhang2024excluir, xu-etal-2025-negconstraint}.
We ask whether an existing retriever can better respect exclusion constraints at inference time, without retraining.

We propose \method, a training-free reranking method that penalizes excluded intent in negative queries with exclusion constraints.
Given a negative query, a large language model produces two short rewrites: a target query for what should be retrieved and a trap query for the excluded intent.
\method penalizes documents similar to the trap intent; retrieval of the excluded-side document is treated as a violation.

Our contributions are as follows.
\begin{itemize}
    \item We introduce \method, a training-free excluded-intent penalty that changes only inference-time scores and requires no retriever retraining.
    \item We formulate exclusion-aware evidence retrieval as a recall--violation trade-off, where systems must retrieve relevant evidence while avoiding evidence the user intended to exclude.
    \item We analyze when document-level trap penalties succeed or fail, identifying cases where relevant documents contain incidental trap evidence or lie close to trap documents in embedding space.
\end{itemize}

\section{Related Work}

\paragraph{Negative constraints in retrieval.}
Language models can be insensitive to negation and struggle to reason under negated statements \citep{truong-etal-2023-language}.
This weakness also appears in retrieval: NevIR tests minimally different negated pairs \citep{weller-etal-2024-nevir}, ExcluIR studies exclusion constraints with answer and trap documents \citep{zhang2024excluir}, BoolQuestions probes Boolean logic in dense retrieval \citep{zhang2024boolquestionsdoesdenseretrieval}, and NegConstraint applies neural-symbolic reranking for negative constraints \citep{xu-etal-2025-negconstraint}.
Unlike neural-symbolic reranking approaches that require explicit logical constraint handling, \method relies only on query-side decomposition and continuous retriever scores.
Similar support/contradiction confusion appears in biomedical QA and vision-language retrieval when negation is central to the evidence \citep{sahoo2026negationsemanticdiagnosingdense, alhamoud2025visionlanguagemodelsunderstandnegation}.

\paragraph{Embedding-level approaches to negation.}
DEO optimizes query embeddings at inference time for negation-aware retrieval \citep{lee2026deotrainingfreedirectembedding}; \method instead keeps retriever and document embeddings fixed and applies a score-level trap penalty.
This makes \method applicable when document embeddings are precomputed or served through closed embedding APIs, where direct embedding optimization or retriever updates are impractical.

\paragraph{Rank fusion for reranking.}
RRF combines ranked lists through reciprocal-rank contributions when raw scores are incomparable \citep{cormack2009reciprocal}.
We considered an anti-RRF variant that subtracts trap-query rank evidence, but use dense scores because they provide a smoother penalty signal and allow a continuous recall--violation frontier.

\section{E-SENS}

\method is a retrieval-time reranking method for exclusion queries.
It operates on fixed retriever scores by decomposing the query into target and trap intents, then lowering the rank of documents similar to the trap query.
The main variant uses the original query as the positive signal and the trap rewrite as a penalty signal; a target-minus-trap variant is reported in the appendix.

\begin{figure*}[t]
    \centering
    \includegraphics[width=0.96\textwidth]{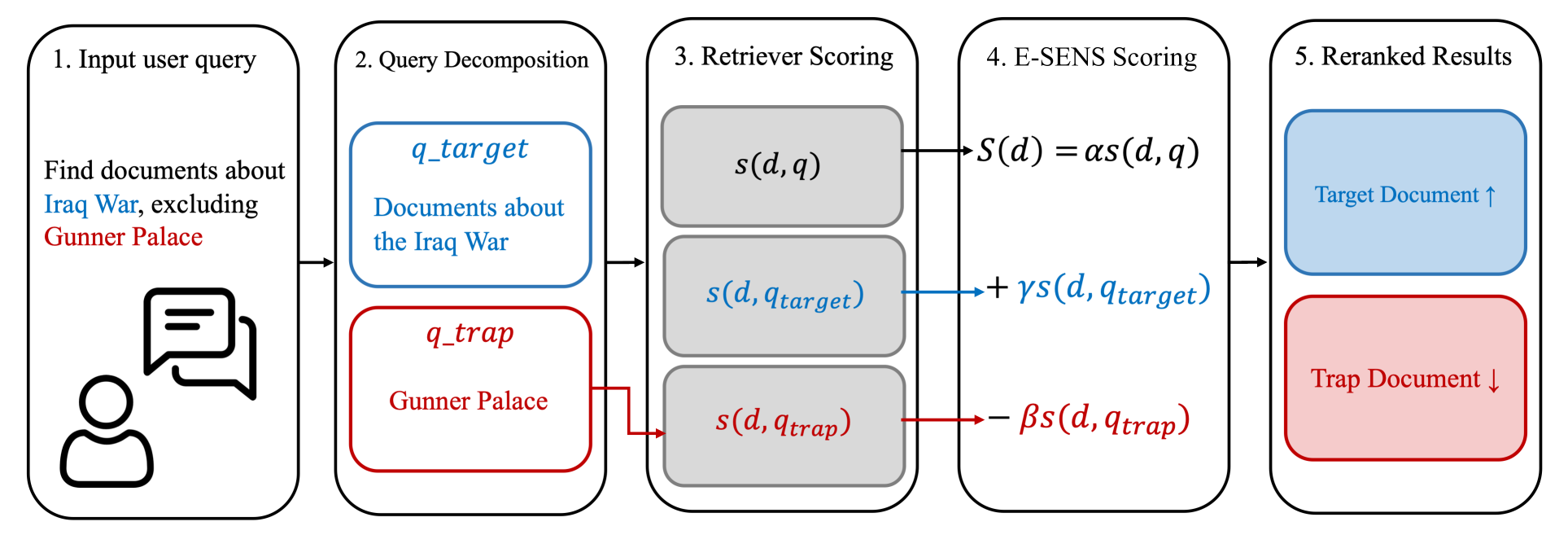}
    \caption{Overview of \method. A negative query with an exclusion constraint is decomposed into target and trap intents. The main score uses the original query as the positive signal and subtracts trap-query similarity. The target-query path is optional and is evaluated as an appendix variant.}
    \label{fig:esens-pipeline}
\end{figure*}

\subsection{Query Decomposition}

Let $q$ be the original user- or dataset-provided query that contains an exclusion constraint.
\method uses GPT-4o mini to produce $q_{\mathrm{target}}$, a positive query with the exclusion removed, and $q_{\mathrm{trap}}$, a compact query for the excluded intent.
Although many traps are named entities or titles, $q_{\mathrm{trap}}$ is not limited to entities: it may refer to an excluded concept, attribute, condition, or constraint.
For example, the query ``location of Friday Harbor Airport without bringing up Dayton International Airport'' is decomposed into target intent ``location of Friday Harbor Airport'' and trap intent ``Dayton International Airport''.
The LLM only produces these strings; it does not score documents, inspect the corpus, or use gold labels.
Appendix~\ref{sec:appendix-decomposition-prompt} provides the decomposition prompt.

\subsection{Score Transformation}

Let $s(d,q)$ be the retriever similarity score between document $d$ and query $q$.
All dense scores use L2-normalized embeddings, so dot product is cosine similarity; for each query rewrite, scores are min-max normalized over the scored document set before combination.
In the main \method variant, we retain the original exclusion query as the positive retrieval signal and use only the trap rewrite as a penalty signal:
\begin{equation}
    S_{\mathrm{main}}(d) = s(d,q) - \beta s(d,q_{\mathrm{trap}}).
    \label{eq:esens-main}
\end{equation}
We use the original query as the main positive signal because target rewrites can remove contextual cues useful for recall.
The target rewrite is used in a target-minus-trap variant, reported in the appendix, to isolate the cost of replacing the positive signal.
More generally, the two scoring variants can be written as
\begin{equation}
    S(d) = \alpha s(d,q) + \gamma s(d,q_{\mathrm{target}})
           - \beta s(d,q_{\mathrm{trap}}),
    \label{eq:esens-general}
\end{equation}
where baseline-minus-trap uses $(\alpha,\gamma)=(1,0)$ and target-minus-trap uses $(0,1)$.
The penalty weight $\beta$ controls how strongly trap-similar documents are demoted.
Table~\ref{tab:main-results} reports fixed beta checkpoints for the main variant, and the appendix reports both scoring variants across beta values.
This formulation requires scores to be comparable across query rewrites; in this paper, we evaluate it with dense retriever scores.
At inference time, \method can rerank either the full corpus or a candidate set: it computes original-query and trap-query scores for the same documents, then applies the transformed score.
Thus, it leaves document embeddings and the retrieval index unchanged; $q_{\mathrm{trap}}$ is the only additional retrieval signal.

\section{Experimental Setup}

\paragraph{Dataset.}
We evaluate on 3,452 negative queries with exclusion constraints from ExcluIR \citep{zhang2024excluir}.
Each example contains an original exclusion query, a gold answer document, and a trap document corresponding to the excluded intent.
We score against the full 90,406-document corpus; the original query is $q$, the answer-side document is $d_{\mathrm{target}}$, and the excluded-side document is $d_{\mathrm{trap}}$.
The decomposer generates $q_{\mathrm{target}}$ and $q_{\mathrm{trap}}$ only from $q$; gold document indices are used only for evaluation.
We use ExcluIR because its document-level target/trap annotations make recall and violation directly measurable, and report a smaller BoolQuestions-not set in the appendix.
Appendix~\ref{sec:appendix-dataset-mapping} details the mapping from dataset fields to $q_{\mathrm{target}}$, $q_{\mathrm{trap}}$, $d_{\mathrm{target}}$, and $d_{\mathrm{trap}}$.

\paragraph{Retrievers and Decomposition.}
We test four embedding models: OpenAI text-embedding-3-small, OpenAI text-embedding-3-large, Qwen3-Embedding-0.6B, and Qwen3-Embedding-4B \citep{zhang2025qwen3embeddingadvancingtext}.
For the main variant, we use the original query as the positive retrieval signal and $q_{\mathrm{trap}}$ only as a penalty signal: $S(d)=s(d,q)-\beta s(d,q_{\mathrm{trap}})$.
We vary $\beta \in [0,1]$ at intervals of 0.01 to characterize the recall--violation trade-off rather than to claim a deployment-ready tuned hyperparameter.
Table~\ref{tab:main-results} reports fixed checkpoints $\beta \in \{0.0,0.1,0.2,0.3\}$, shared across all retrievers rather than selected per retriever.
Appendix Figures~\ref{fig:appendix-excluir-beta-curves}--\ref{fig:appendix-boolquestions-beta-curves} plot the full sweep.

\paragraph{Metrics.}
Recall@$k$ measures whether the gold answer document appears in the top $k$; Violation@$k$ measures whether the trap document appears in the top $k$.
Higher recall and lower violation are better.
We report R@5 and V@5 in the main table as representative top-$k$ metrics, while Avg. R and Avg. V average over $k \in \{3,5,7,9\}$ to summarize the overall trade-off without selecting a favorable $k$.

\section{Results}

\begin{table*}[t]
\centering
\small
\setlength{\tabcolsep}{4.2pt}
\begin{tabular}{@{}lrrrrrrr@{}}
\toprule
Retriever & $\beta$ & R@5$\uparrow$ & V@5$\downarrow$ & Avg. R$\uparrow$ & $\Delta$Avg. R & Avg. V$\downarrow$ & $\Delta$Avg. V$\downarrow$ \\
\midrule
text-embedding-3-small & 0.00 & \textbf{0.9377} & 0.8789 & 0.9366 & -- & 0.8788 & -- \\
 & 0.10 & \textbf{0.9377} & 0.7961 & \textbf{0.9379} & \textbf{+0.0014} & 0.8001 & -0.0787 \\
 & 0.20 & \textbf{0.9377} & 0.6753 & 0.9375 & +0.0009 & 0.6842 & -0.1946 \\
 & 0.30 & 0.9366 & \textbf{0.5426} & 0.9369 & +0.0004 & \textbf{0.5502} & \textbf{-0.3286} \\
\midrule
text-embedding-3-large & 0.00 & 0.9400 & 0.8862 & 0.9412 & -- & 0.8871 & -- \\
 & 0.10 & \textbf{0.9415} & 0.8108 & \textbf{0.9424} & \textbf{+0.0012} & 0.8132 & -0.0739 \\
 & 0.20 & 0.9392 & 0.6889 & 0.9408 & -0.0004 & 0.6943 & -0.1928 \\
 & 0.30 & 0.9389 & \textbf{0.5437} & 0.9391 & -0.0021 & \textbf{0.5537} & \textbf{-0.3334} \\
\midrule
Qwen3-Embedding-0.6B & 0.00 & \textbf{0.9360} & 0.7277 & \textbf{0.9383} & -- & 0.7320 & -- \\
 & 0.10 & 0.9354 & 0.6202 & 0.9374 & -0.0009 & 0.6262 & -0.1058 \\
 & 0.20 & 0.9334 & 0.5061 & 0.9345 & -0.0038 & 0.5147 & -0.2173 \\
 & 0.30 & 0.9305 & \textbf{0.3931} & 0.9312 & -0.0071 & \textbf{0.4007} & \textbf{-0.3313} \\
\midrule
Qwen3-Embedding-4B & 0.00 & \textbf{0.9421} & 0.7361 & 0.9422 & -- & 0.7412 & -- \\
 & 0.10 & \textbf{0.9421} & 0.6446 & \textbf{0.9424} & \textbf{+0.0002} & 0.6485 & -0.0928 \\
 & 0.20 & 0.9403 & 0.5411 & 0.9402 & -0.0020 & 0.5456 & -0.1957 \\
 & 0.30 & 0.9377 & \textbf{0.4348} & 0.9393 & -0.0029 & \textbf{0.4438} & \textbf{-0.2974} \\
\bottomrule
\end{tabular}
\caption{Fixed beta checkpoints for the main baseline-minus-trap variant on ExcluIR. R is recall, V is trap violation, Avg. averages over $k \in \{3,5,7,9\}$, and deltas are relative to $\beta=0.00$ for the same retriever. Bold marks the best value within each retriever block; ties are bolded.}
\label{tab:main-results}
\end{table*}

Table~\ref{tab:main-results} summarizes the recall--violation trade-off for \method across fixed beta checkpoints.
The $\beta=0.00$ rows correspond to the original retriever scores, which achieve high recall but often retrieve the trap document as well, with violation@5 ranging from 0.728 to 0.886.
As $\beta$ increases, trap retrieval decreases consistently across retrievers while recall remains close to the baseline over moderate penalty strengths, showing that the gains are not confined to a single tuned beta.
At $\beta=0.10$, average recall changes by only $-0.0009$ to $+0.0014$, while average violation drops by 0.074--0.106; at $\beta=0.20$, average violation drops by 0.193--0.217 with at most 0.0038 recall loss.
The $\beta=0.30$ checkpoint further reduces average violation by 0.297--0.333, with a larger but still modest recall decrease for Qwen3-Embedding-0.6B.
At $\beta=1.0$, average violation drops to 0.030--0.036, but average recall falls to 0.779--0.835, illustrating the cost of over-penalizing the trap intent.
Appendix curves show the full frontier for both scoring variants and make the same pattern visible across the complete sweep.

Appendix Tables~\ref{tab:appendix-excluir-target-text-embedding-3-small-beta-grid}--\ref{tab:appendix-excluir-target-qwen3-4b-beta-grid} clarify why we use the original query, rather than the target rewrite, as the main positive signal.
At $\beta=0$, target-minus-trap already lowers violation substantially, but with lower recall than the original-query baseline, suggesting that target rewrites can discard answer-retrieval context.
The main baseline-minus-trap variant is therefore conservative: it preserves the original retrieval signal and uses the trap query only to demote constraint-violating documents.

\section{Failure Analysis}

We manually inspected cases where the trap penalty affects answer ranks and found two recurring patterns.
Some target documents mention the excluded entity incidentally, such as a film page listing an excluded actor, while other answer and trap documents are close embedding neighbors, such as sibling entities or adaptations.
These failures reflect document-level granularity: a valid answer may briefly mention the excluded side, whereas a trap document is primarily about it.

\section{Conclusion}

We presented \method, a training-free excluded-intent penalty for negative queries with exclusion constraints.
By extracting a compact excluded-side query and subtracting its similarity score, \method turns exclusion into a score-level penalty that can be applied without retraining or changing the corpus.
Across retrievers, moderate $\beta$ values preserve answer recall while reducing trap retrieval, showing that negative constraints are better treated as retrieval-time selection constraints than only as surface negation in the query.
This matters in retrieval-augmented generation because a high-ranked trap document can steer the generator toward excluded evidence, and the consistent pattern across four embedding models suggests a shared dense-retrieval failure.
The recall--violation frontier offers a practical way to choose $\beta$: lower values favor recall, while higher values more aggressively suppress excluded-side documents.
Future work can extend \method with passage- or span-level verification to distinguish incidental mentions from truly constraint-violating evidence.

\section*{Limitations}

\method adds inference cost because it requires query decomposition, an additional trap-query score, and reranking.
It assumes that the decomposition reliably isolates the excluded side and that scores from different query rewrites remain comparable after normalization.
Our experiments tune $\beta$ on validation data and do not report statistical uncertainty estimates, so small recall differences should be interpreted cautiously.
Finally, the document-level penalty cannot distinguish a valid answer that only mentions the excluded entity incidentally from a document that is mainly about the excluded side.

\clearpage
\bibliography{custom}

\clearpage
\appendix

\section{Experimental Details}

\paragraph{Compute.}
All experiments were run on a single RTX 4090 GPU with 16 vCPUs (AMD Ryzen 9 7950X 16-Core Processor), 61 GB memory, and a 20 GB container disk.

\section{Query Decomposition Prompt}
\label{sec:appendix-decomposition-prompt}

The template below was used with GPT-4o mini at temperature 0 to generate
$q_{\mathrm{target}}$ and $q_{\mathrm{trap}}$.

\begingroup
\small
\setlength{\parindent}{0pt}
\setlength{\parskip}{0.45em}
\textbf{Query Decomposition Prompt Template}

\textbf{Role:}
You are a recall-preserving query decomposition module for
exclusion-aware information retrieval.

\textbf{Task:}
Given only the original query, produce two retrieval queries:
$q_{\mathrm{target}}$, the positive information need, and
$q_{\mathrm{trap}}$, a short high-precision anchor for the unwanted,
excluded, contrastive, or constraint-violating side.
Do not use gold documents, trap documents, answers, relevance labels,
dataset annotations, or external knowledge beyond what is directly implied
by the query.

\textbf{Decomposition Rules:}
Preserve the main search intent in $q_{\mathrm{target}}$ and remove only
phrases that point to the unwanted side.
$q_{\mathrm{target}}$ should remain independently useful for retrieval and
should not encode the exclusion with wrappers such as ``not'', ``without'',
``excluding'', ``except'', ``instead of'', or ``rather than'', unless the
negated phrase is the natural name of the desired concept.
$q_{\mathrm{trap}}$ should be the strongest compact retrieval anchor for
the unwanted side: preferably an explicitly excluded named entity, title,
organization, person, work, place, product, event, date, object, attribute,
condition, relation, role, or category.
Strip exclusion wrappers from $q_{\mathrm{trap}}$ and avoid broad anchors
such as ``career'', ``film'', ``country'', or ``person'' unless the unwanted
side is exactly that broad.
If the query contains a direct contrast, use the obvious counterpart when
available (e.g., legal $\leftrightarrow$ illegal, allowed
$\leftrightarrow$ forbidden, paid $\leftrightarrow$ unpaid, present
$\leftrightarrow$ absent, won $\leftrightarrow$ lost).
If no direct counterpart is obvious, use the explicit unwanted phrase.
Set $q_{\mathrm{trap}}$ to an empty string only when no excluded,
negative, contrastive, exception, or undesired side is present.

\textbf{Output Format:}
Return only valid JSON with exactly two fields:

\noindent{\ttfamily\scriptsize
\{\\
\hspace*{1em}"q\_target": "...",\\
\hspace*{1em}"q\_trap": "..."\\
\}
}

\textbf{User Message Template:}

\noindent{\ttfamily\scriptsize
Now decompose the following query.\\[0.2em]
Original query:\\
\{original\_query\}\\[0.2em]
Return only valid JSON. Do not include markdown, comments,
explanations, or extra fields.\\[0.2em]
\{\\
\hspace*{1em}"q\_target": "...",\\
\hspace*{1em}"q\_trap": "..."\\
\}
}
\endgroup

\section{Dataset Mapping and BoolQuestions-not Set}
\label{sec:appendix-dataset-mapping}

ExcluIR provides a natural four-part structure for our setting.
The original exclusion query is $q$, the answer-side corpus index is $d_{\mathrm{target}}$, and the excluded-side corpus index is $d_{\mathrm{trap}}$.
Its paired construction also makes oracle-style query strings recoverable: the original positive question can be viewed as $q_{\mathrm{target}}$, and the title or short identifier of the excluded document can be viewed as $q_{\mathrm{trap}}$.
In the reported \method experiments, however, $q_{\mathrm{target}}$ and $q_{\mathrm{trap}}$ are generated only from $q$ by GPT-4o mini; gold document indices are used only for evaluation.

For other negative-constraint retrieval datasets, $q_{\mathrm{target}}$ is the positive information need after removing the exclusion, $q_{\mathrm{trap}}$ is the compact excluded-side query, $d_{\mathrm{target}}$ is a document satisfying the positive need, and $d_{\mathrm{trap}}$ is a plausible but constraint-violating document.
When a dataset provides positive and negative passages, we map them to $d_{\mathrm{target}}$ and $d_{\mathrm{trap}}$ respectively.
When it provides only documents, $q_{\mathrm{target}}$ and $q_{\mathrm{trap}}$ can be produced by the same query-only decomposer.

As an additional check, we evaluate the subset of BoolQuestions \citep{zhang2024boolquestionsdoesdenseretrieval} with \texttt{question\_type = not}.
Each example includes a positive passage and a negative passage that can be treated as $d_{\mathrm{target}}$ and $d_{\mathrm{trap}}$.
This yields 323 examples, 195 from MS MARCO and 128 from Natural Questions, over a 646-document evaluation corpus.
Because BoolQuestions does not provide explicit $q_{\mathrm{target}}$ and $q_{\mathrm{trap}}$ fields, we generate them with the same GPT-4o mini decomposer and report results across penalty-weight settings in Figure~\ref{fig:appendix-boolquestions-beta-curves} and Tables~\ref{tab:appendix-boolquestions-baseline-text-embedding-3-small-beta-grid}--\ref{tab:appendix-boolquestions-target-qwen3-4b-beta-grid}.

The full $\beta$ results are included to show the recall--violation trade-off across penalty strengths rather than to imply that $\beta$ should be selected on the test set.

\section{Additional Results}
\label{sec:appendix-additional-results}
\begin{figure*}[p]
\centering
\includegraphics[width=0.98\textwidth]{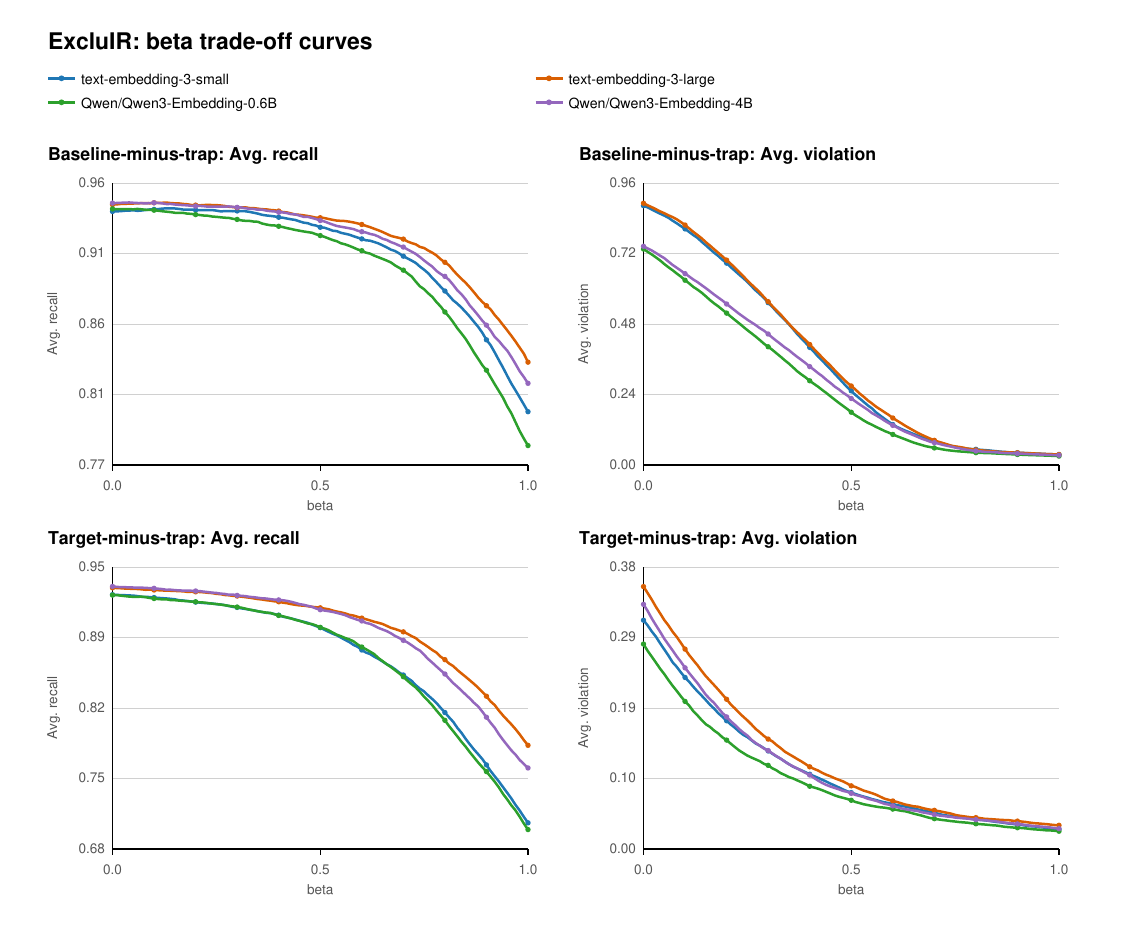}
\caption{ExcluIR recall--violation trade-off curves over the full beta sweep. The top row uses the baseline-minus-trap score, and the bottom row uses the target-minus-trap score.}
\label{fig:appendix-excluir-beta-curves}
\end{figure*}

\begin{figure*}[p]
\centering
\includegraphics[width=0.98\textwidth]{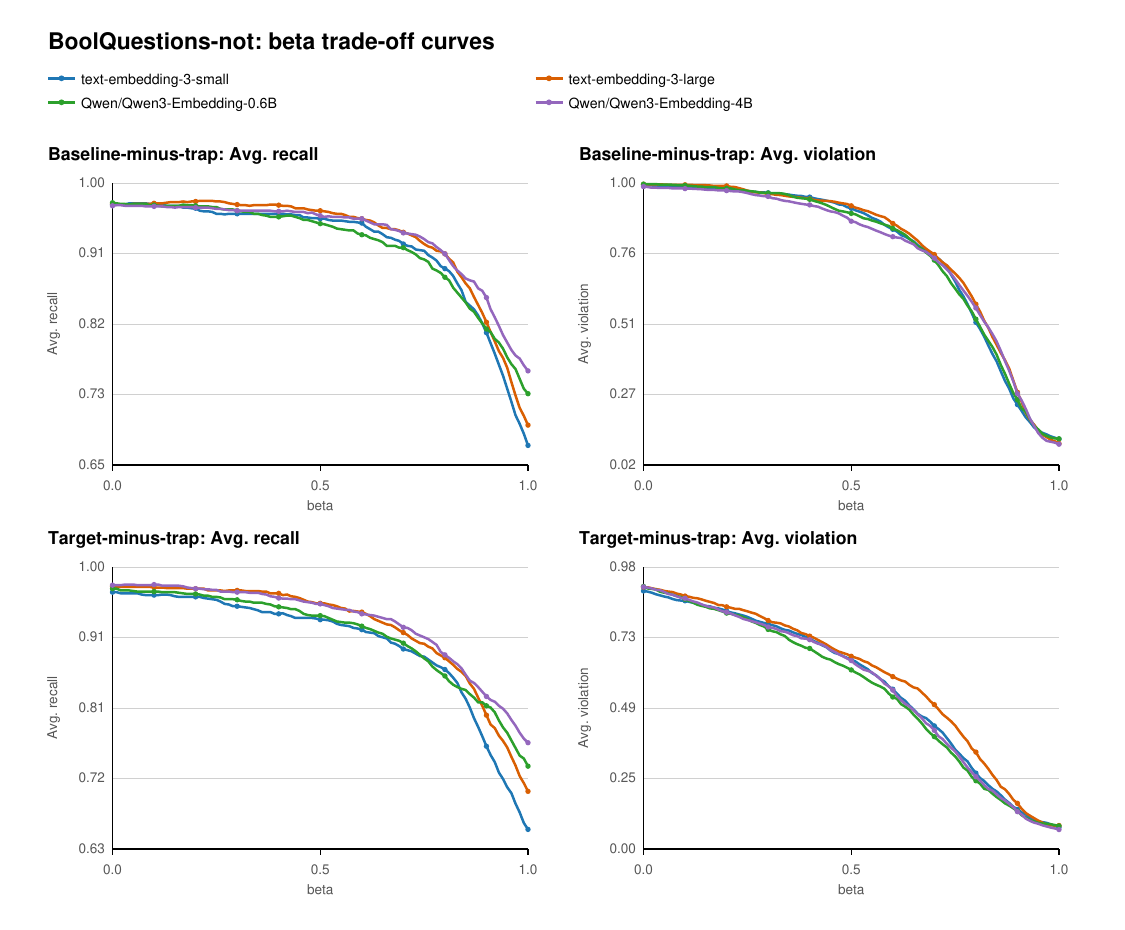}
\caption{BoolQuestions-not recall--violation trade-off curves over the full beta sweep. The top row uses the baseline-minus-trap score, and the bottom row uses the target-minus-trap score.}
\label{fig:appendix-boolquestions-beta-curves}
\end{figure*}

\noindent Figures~\ref{fig:appendix-excluir-beta-curves} and~\ref{fig:appendix-boolquestions-beta-curves} plot the full $\beta=0.00$--$1.00$ sweep at 0.01 intervals. Tables~\ref{tab:appendix-excluir-baseline-text-embedding-3-small-beta-grid}--\ref{tab:appendix-boolquestions-target-qwen3-4b-beta-grid} report 0.10 beta checkpoints in portrait tables, split by dataset, scoring variant, and retriever for readability.

\subsection{Beta Trade-off Curves}

\clearpage
\makeatletter
\setlength{\@dblfptop}{0pt}
\setlength{\@dblfpsep}{10pt plus 1pt minus 1pt}
\setlength{\@dblfpbot}{0pt plus 1fil}
\makeatother
\begin{table*}[t]
\centering
\begingroup
\small
\setlength{\tabcolsep}{4.2pt}
\renewcommand{\arraystretch}{1.05}
\begin{tabular}{@{}r*{8}{c}@{}}
\toprule
$\beta$ & \multicolumn{2}{c}{$k=3$} & \multicolumn{2}{c}{$k=5$} & \multicolumn{2}{c}{$k=7$} & \multicolumn{2}{c}{$k=9$} \\
\cmidrule(lr){2-3} \cmidrule(lr){4-5} \cmidrule(lr){6-7} \cmidrule(lr){8-9}
 & R & V & R & V & R & V & R & V \\
\midrule
0.0 & 0.9166 & 0.8250 & 0.9377 & 0.8789 & 0.9447 & 0.8992 & 0.9473 & 0.9119 \\
0.1 & 0.9206 & 0.7228 & 0.9377 & 0.7961 & 0.9441 & 0.8308 & 0.9493 & 0.8508 \\
0.2 & 0.9224 & 0.5982 & 0.9377 & 0.6753 & 0.9426 & 0.7176 & 0.9473 & 0.7457 \\
0.3 & 0.9241 & 0.4612 & 0.9366 & 0.5426 & 0.9409 & 0.5817 & 0.9461 & 0.6153 \\
0.4 & 0.9195 & 0.3140 & 0.9308 & 0.3885 & 0.9383 & 0.4311 & 0.9421 & 0.4603 \\
0.5 & 0.9134 & 0.1819 & 0.9250 & 0.2390 & 0.9296 & 0.2787 & 0.9357 & 0.3059 \\
0.6 & 0.9073 & 0.1034 & 0.9163 & 0.1283 & 0.9221 & 0.1501 & 0.9267 & 0.1669 \\
0.7 & 0.8946 & 0.0663 & 0.9050 & 0.0788 & 0.9108 & 0.0869 & 0.9154 & 0.0913 \\
0.8 & 0.8688 & 0.0463 & 0.8801 & 0.0527 & 0.8867 & 0.0559 & 0.8963 & 0.0576 \\
0.9 & 0.8291 & 0.0371 & 0.8476 & 0.0414 & 0.8592 & 0.0429 & 0.8644 & 0.0455 \\
1.0 & 0.7743 & 0.0313 & 0.7987 & 0.0362 & 0.8134 & 0.0371 & 0.8201 & 0.0379 \\
\bottomrule
\end{tabular}
\caption{ExcluIR beta checkpoint results for text-embedding-3-small with the baseline-minus-trap score. Columns report recall (R) and trap violation (V) at $k \in \{3,5,7,9\}$; rows use 0.10 beta checkpoints. The score is $S(d)=s(d,q)-\beta s(d,q_{\mathrm{trap}})$.}
\label{tab:appendix-excluir-baseline-text-embedding-3-small-beta-grid}
\endgroup
\end{table*}

\begin{table*}[t]
\centering
\begingroup
\small
\setlength{\tabcolsep}{4.2pt}
\renewcommand{\arraystretch}{1.05}
\begin{tabular}{@{}r*{8}{c}@{}}
\toprule
$\beta$ & \multicolumn{2}{c}{$k=3$} & \multicolumn{2}{c}{$k=5$} & \multicolumn{2}{c}{$k=7$} & \multicolumn{2}{c}{$k=9$} \\
\cmidrule(lr){2-3} \cmidrule(lr){4-5} \cmidrule(lr){6-7} \cmidrule(lr){8-9}
 & R & V & R & V & R & V & R & V \\
\midrule
0.0 & 0.9270 & 0.8323 & 0.9400 & 0.8862 & 0.9467 & 0.9082 & 0.9510 & 0.9218 \\
0.1 & 0.9284 & 0.7396 & 0.9415 & 0.8108 & 0.9476 & 0.8415 & 0.9522 & 0.8607 \\
0.2 & 0.9273 & 0.5999 & 0.9392 & 0.6889 & 0.9452 & 0.7294 & 0.9516 & 0.7590 \\
0.3 & 0.9256 & 0.4568 & 0.9389 & 0.5437 & 0.9438 & 0.5921 & 0.9481 & 0.6220 \\
0.4 & 0.9238 & 0.3224 & 0.9360 & 0.3972 & 0.9418 & 0.4421 & 0.9458 & 0.4719 \\
0.5 & 0.9195 & 0.1973 & 0.9311 & 0.2590 & 0.9380 & 0.2955 & 0.9409 & 0.3195 \\
0.6 & 0.9148 & 0.1156 & 0.9276 & 0.1506 & 0.9319 & 0.1758 & 0.9368 & 0.1950 \\
0.7 & 0.9073 & 0.0658 & 0.9157 & 0.0785 & 0.9218 & 0.0895 & 0.9270 & 0.1005 \\
0.8 & 0.8864 & 0.0437 & 0.9006 & 0.0504 & 0.9087 & 0.0545 & 0.9134 & 0.0559 \\
0.9 & 0.8569 & 0.0368 & 0.8705 & 0.0411 & 0.8801 & 0.0432 & 0.8847 & 0.0452 \\
1.0 & 0.8129 & 0.0316 & 0.8308 & 0.0345 & 0.8439 & 0.0374 & 0.8525 & 0.0385 \\
\bottomrule
\end{tabular}
\caption{ExcluIR beta checkpoint results for text-embedding-3-large with the baseline-minus-trap score. Columns report recall (R) and trap violation (V) at $k \in \{3,5,7,9\}$; rows use 0.10 beta checkpoints. The score is $S(d)=s(d,q)-\beta s(d,q_{\mathrm{trap}})$.}
\label{tab:appendix-excluir-baseline-text-embedding-3-large-beta-grid}
\endgroup
\end{table*}

\begin{table*}[t]
\centering
\begingroup
\small
\setlength{\tabcolsep}{4.2pt}
\renewcommand{\arraystretch}{1.05}
\begin{tabular}{@{}r*{8}{c}@{}}
\toprule
$\beta$ & \multicolumn{2}{c}{$k=3$} & \multicolumn{2}{c}{$k=5$} & \multicolumn{2}{c}{$k=7$} & \multicolumn{2}{c}{$k=9$} \\
\cmidrule(lr){2-3} \cmidrule(lr){4-5} \cmidrule(lr){6-7} \cmidrule(lr){8-9}
 & R & V & R & V & R & V & R & V \\
\midrule
0.0 & 0.9238 & 0.6654 & 0.9360 & 0.7277 & 0.9450 & 0.7598 & 0.9484 & 0.7749 \\
0.1 & 0.9238 & 0.5556 & 0.9354 & 0.6202 & 0.9426 & 0.6541 & 0.9479 & 0.6747 \\
0.2 & 0.9203 & 0.4458 & 0.9334 & 0.5061 & 0.9395 & 0.5420 & 0.9447 & 0.5649 \\
0.3 & 0.9166 & 0.3346 & 0.9305 & 0.3931 & 0.9363 & 0.4293 & 0.9415 & 0.4458 \\
0.4 & 0.9145 & 0.2306 & 0.9250 & 0.2798 & 0.9316 & 0.3045 & 0.9351 & 0.3268 \\
0.5 & 0.9082 & 0.1350 & 0.9189 & 0.1712 & 0.9250 & 0.1952 & 0.9293 & 0.2141 \\
0.6 & 0.8966 & 0.0797 & 0.9093 & 0.0988 & 0.9160 & 0.1127 & 0.9186 & 0.1228 \\
0.7 & 0.8786 & 0.0484 & 0.8948 & 0.0565 & 0.9050 & 0.0614 & 0.9096 & 0.0649 \\
0.8 & 0.8470 & 0.0374 & 0.8670 & 0.0411 & 0.8772 & 0.0435 & 0.8841 & 0.0455 \\
0.9 & 0.8048 & 0.0319 & 0.8259 & 0.0351 & 0.8389 & 0.0371 & 0.8488 & 0.0382 \\
1.0 & 0.7500 & 0.0258 & 0.7758 & 0.0298 & 0.7906 & 0.0322 & 0.7990 & 0.0333 \\
\bottomrule
\end{tabular}
\caption{ExcluIR beta checkpoint results for Qwen/Qwen3-Embedding-0.6B with the baseline-minus-trap score. Columns report recall (R) and trap violation (V) at $k \in \{3,5,7,9\}$; rows use 0.10 beta checkpoints. The score is $S(d)=s(d,q)-\beta s(d,q_{\mathrm{trap}})$.}
\label{tab:appendix-excluir-baseline-qwen3-0p6b-beta-grid}
\endgroup
\end{table*}

\begin{table*}[t]
\centering
\begingroup
\small
\setlength{\tabcolsep}{4.2pt}
\renewcommand{\arraystretch}{1.05}
\begin{tabular}{@{}r*{8}{c}@{}}
\toprule
$\beta$ & \multicolumn{2}{c}{$k=3$} & \multicolumn{2}{c}{$k=5$} & \multicolumn{2}{c}{$k=7$} & \multicolumn{2}{c}{$k=9$} \\
\cmidrule(lr){2-3} \cmidrule(lr){4-5} \cmidrule(lr){6-7} \cmidrule(lr){8-9}
 & R & V & R & V & R & V & R & V \\
\midrule
0.0 & 0.9282 & 0.6816 & 0.9421 & 0.7361 & 0.9467 & 0.7665 & 0.9519 & 0.7807 \\
0.1 & 0.9296 & 0.5834 & 0.9421 & 0.6446 & 0.9476 & 0.6732 & 0.9505 & 0.6926 \\
0.2 & 0.9267 & 0.4780 & 0.9403 & 0.5411 & 0.9447 & 0.5724 & 0.9490 & 0.5907 \\
0.3 & 0.9270 & 0.3780 & 0.9377 & 0.4348 & 0.9429 & 0.4710 & 0.9496 & 0.4913 \\
0.4 & 0.9247 & 0.2732 & 0.9337 & 0.3276 & 0.9403 & 0.3566 & 0.9455 & 0.3763 \\
0.5 & 0.9189 & 0.1758 & 0.9273 & 0.2173 & 0.9348 & 0.2436 & 0.9409 & 0.2648 \\
0.6 & 0.9105 & 0.1002 & 0.9203 & 0.1283 & 0.9282 & 0.1469 & 0.9328 & 0.1614 \\
0.7 & 0.8980 & 0.0594 & 0.9099 & 0.0727 & 0.9186 & 0.0831 & 0.9238 & 0.0875 \\
0.8 & 0.8769 & 0.0406 & 0.8919 & 0.0458 & 0.8989 & 0.0504 & 0.9035 & 0.0539 \\
0.9 & 0.8418 & 0.0333 & 0.8578 & 0.0382 & 0.8670 & 0.0408 & 0.8731 & 0.0426 \\
1.0 & 0.7992 & 0.0287 & 0.8175 & 0.0319 & 0.8288 & 0.0348 & 0.8375 & 0.0371 \\
\bottomrule
\end{tabular}
\caption{ExcluIR beta checkpoint results for Qwen/Qwen3-Embedding-4B with the baseline-minus-trap score. Columns report recall (R) and trap violation (V) at $k \in \{3,5,7,9\}$; rows use 0.10 beta checkpoints. The score is $S(d)=s(d,q)-\beta s(d,q_{\mathrm{trap}})$.}
\label{tab:appendix-excluir-baseline-qwen3-4b-beta-grid}
\endgroup
\end{table*}

\begin{table*}[t]
\centering
\begingroup
\small
\setlength{\tabcolsep}{4.2pt}
\renewcommand{\arraystretch}{1.05}
\begin{tabular}{@{}r*{8}{c}@{}}
\toprule
$\beta$ & \multicolumn{2}{c}{$k=3$} & \multicolumn{2}{c}{$k=5$} & \multicolumn{2}{c}{$k=7$} & \multicolumn{2}{c}{$k=9$} \\
\cmidrule(lr){2-3} \cmidrule(lr){4-5} \cmidrule(lr){6-7} \cmidrule(lr){8-9}
 & R & V & R & V & R & V & R & V \\
\midrule
0.0 & 0.9148 & 0.2422 & 0.9267 & 0.3013 & 0.9340 & 0.3340 & 0.9368 & 0.3598 \\
0.1 & 0.9122 & 0.1813 & 0.9241 & 0.2257 & 0.9299 & 0.2509 & 0.9351 & 0.2697 \\
0.2 & 0.9061 & 0.1370 & 0.9209 & 0.1689 & 0.9267 & 0.1877 & 0.9302 & 0.1999 \\
0.3 & 0.9012 & 0.1054 & 0.9154 & 0.1301 & 0.9215 & 0.1428 & 0.9256 & 0.1521 \\
0.4 & 0.8922 & 0.0826 & 0.9082 & 0.0973 & 0.9145 & 0.1092 & 0.9189 & 0.1153 \\
0.5 & 0.8775 & 0.0634 & 0.8946 & 0.0750 & 0.9035 & 0.0808 & 0.9105 & 0.0855 \\
0.6 & 0.8520 & 0.0516 & 0.8728 & 0.0603 & 0.8830 & 0.0637 & 0.8925 & 0.0672 \\
0.7 & 0.8265 & 0.0420 & 0.8468 & 0.0487 & 0.8612 & 0.0524 & 0.8699 & 0.0530 \\
0.8 & 0.7853 & 0.0345 & 0.8120 & 0.0397 & 0.8256 & 0.0423 & 0.8375 & 0.0437 \\
0.9 & 0.7303 & 0.0278 & 0.7616 & 0.0327 & 0.7775 & 0.0348 & 0.7903 & 0.0362 \\
1.0 & 0.6698 & 0.0229 & 0.7063 & 0.0261 & 0.7245 & 0.0278 & 0.7370 & 0.0295 \\
\bottomrule
\end{tabular}
\caption{ExcluIR beta checkpoint results for text-embedding-3-small with the target-minus-trap score. Columns report recall (R) and trap violation (V) at $k \in \{3,5,7,9\}$; rows use 0.10 beta checkpoints. The score is $S(d)=s(d,q_{\mathrm{target}})-\beta s(d,q_{\mathrm{trap}})$.}
\label{tab:appendix-excluir-target-text-embedding-3-small-beta-grid}
\endgroup
\end{table*}

\begin{table*}[t]
\centering
\begingroup
\small
\setlength{\tabcolsep}{4.2pt}
\renewcommand{\arraystretch}{1.05}
\begin{tabular}{@{}r*{8}{c}@{}}
\toprule
$\beta$ & \multicolumn{2}{c}{$k=3$} & \multicolumn{2}{c}{$k=5$} & \multicolumn{2}{c}{$k=7$} & \multicolumn{2}{c}{$k=9$} \\
\cmidrule(lr){2-3} \cmidrule(lr){4-5} \cmidrule(lr){6-7} \cmidrule(lr){8-9}
 & R & V & R & V & R & V & R & V \\
\midrule
0.0 & 0.9203 & 0.2816 & 0.9325 & 0.3462 & 0.9406 & 0.3815 & 0.9452 & 0.4099 \\
0.1 & 0.9183 & 0.2103 & 0.9299 & 0.2622 & 0.9392 & 0.2908 & 0.9429 & 0.3178 \\
0.2 & 0.9189 & 0.1567 & 0.9290 & 0.1932 & 0.9357 & 0.2196 & 0.9397 & 0.2402 \\
0.3 & 0.9140 & 0.1170 & 0.9258 & 0.1437 & 0.9316 & 0.1608 & 0.9357 & 0.1729 \\
0.4 & 0.9085 & 0.0924 & 0.9198 & 0.1075 & 0.9270 & 0.1191 & 0.9299 & 0.1254 \\
0.5 & 0.9038 & 0.0718 & 0.9134 & 0.0831 & 0.9203 & 0.0901 & 0.9244 & 0.0965 \\
0.6 & 0.8902 & 0.0548 & 0.9053 & 0.0640 & 0.9119 & 0.0681 & 0.9151 & 0.0727 \\
0.7 & 0.8740 & 0.0461 & 0.8914 & 0.0519 & 0.8983 & 0.0548 & 0.9061 & 0.0565 \\
0.8 & 0.8456 & 0.0379 & 0.8627 & 0.0417 & 0.8734 & 0.0446 & 0.8815 & 0.0455 \\
0.9 & 0.8065 & 0.0339 & 0.8276 & 0.0371 & 0.8389 & 0.0391 & 0.8499 & 0.0408 \\
1.0 & 0.7590 & 0.0293 & 0.7801 & 0.0316 & 0.7940 & 0.0333 & 0.8016 & 0.0339 \\
\bottomrule
\end{tabular}
\caption{ExcluIR beta checkpoint results for text-embedding-3-large with the target-minus-trap score. Columns report recall (R) and trap violation (V) at $k \in \{3,5,7,9\}$; rows use 0.10 beta checkpoints. The score is $S(d)=s(d,q_{\mathrm{target}})-\beta s(d,q_{\mathrm{trap}})$.}
\label{tab:appendix-excluir-target-text-embedding-3-large-beta-grid}
\endgroup
\end{table*}

\begin{table*}[t]
\centering
\begingroup
\small
\setlength{\tabcolsep}{4.2pt}
\renewcommand{\arraystretch}{1.05}
\begin{tabular}{@{}r*{8}{c}@{}}
\toprule
$\beta$ & \multicolumn{2}{c}{$k=3$} & \multicolumn{2}{c}{$k=5$} & \multicolumn{2}{c}{$k=7$} & \multicolumn{2}{c}{$k=9$} \\
\cmidrule(lr){2-3} \cmidrule(lr){4-5} \cmidrule(lr){6-7} \cmidrule(lr){8-9}
 & R & V & R & V & R & V & R & V \\
\midrule
0.0 & 0.9137 & 0.2173 & 0.9256 & 0.2720 & 0.9340 & 0.3007 & 0.9377 & 0.3184 \\
0.1 & 0.9105 & 0.1587 & 0.9221 & 0.1906 & 0.9302 & 0.2147 & 0.9348 & 0.2341 \\
0.2 & 0.9096 & 0.1208 & 0.9198 & 0.1428 & 0.9256 & 0.1573 & 0.9302 & 0.1680 \\
0.3 & 0.9024 & 0.0944 & 0.9160 & 0.1127 & 0.9215 & 0.1191 & 0.9250 & 0.1263 \\
0.4 & 0.8931 & 0.0713 & 0.9079 & 0.0829 & 0.9140 & 0.0898 & 0.9177 & 0.0953 \\
0.5 & 0.8760 & 0.0565 & 0.8960 & 0.0658 & 0.9044 & 0.0689 & 0.9105 & 0.0724 \\
0.6 & 0.8523 & 0.0455 & 0.8775 & 0.0536 & 0.8876 & 0.0574 & 0.8943 & 0.0594 \\
0.7 & 0.8213 & 0.0359 & 0.8470 & 0.0391 & 0.8595 & 0.0429 & 0.8699 & 0.0461 \\
0.8 & 0.7755 & 0.0307 & 0.8033 & 0.0339 & 0.8207 & 0.0351 & 0.8311 & 0.0368 \\
0.9 & 0.7242 & 0.0252 & 0.7538 & 0.0287 & 0.7729 & 0.0301 & 0.7830 & 0.0310 \\
1.0 & 0.6648 & 0.0214 & 0.6979 & 0.0238 & 0.7181 & 0.0246 & 0.7309 & 0.0264 \\
\bottomrule
\end{tabular}
\caption{ExcluIR beta checkpoint results for Qwen/Qwen3-Embedding-0.6B with the target-minus-trap score. Columns report recall (R) and trap violation (V) at $k \in \{3,5,7,9\}$; rows use 0.10 beta checkpoints. The score is $S(d)=s(d,q_{\mathrm{target}})-\beta s(d,q_{\mathrm{trap}})$.}
\label{tab:appendix-excluir-target-qwen3-0p6b-beta-grid}
\endgroup
\end{table*}

\begin{table*}[t]
\centering
\begingroup
\small
\setlength{\tabcolsep}{4.2pt}
\renewcommand{\arraystretch}{1.05}
\begin{tabular}{@{}r*{8}{c}@{}}
\toprule
$\beta$ & \multicolumn{2}{c}{$k=3$} & \multicolumn{2}{c}{$k=5$} & \multicolumn{2}{c}{$k=7$} & \multicolumn{2}{c}{$k=9$} \\
\cmidrule(lr){2-3} \cmidrule(lr){4-5} \cmidrule(lr){6-7} \cmidrule(lr){8-9}
 & R & V & R & V & R & V & R & V \\
\midrule
0.0 & 0.9232 & 0.2613 & 0.9345 & 0.3213 & 0.9409 & 0.3560 & 0.9452 & 0.3844 \\
0.1 & 0.9221 & 0.1915 & 0.9316 & 0.2361 & 0.9397 & 0.2656 & 0.9432 & 0.2865 \\
0.2 & 0.9192 & 0.1437 & 0.9308 & 0.1715 & 0.9363 & 0.1929 & 0.9403 & 0.2063 \\
0.3 & 0.9143 & 0.1110 & 0.9253 & 0.1292 & 0.9331 & 0.1411 & 0.9368 & 0.1509 \\
0.4 & 0.9102 & 0.0817 & 0.9218 & 0.0976 & 0.9284 & 0.1066 & 0.9319 & 0.1141 \\
0.5 & 0.8980 & 0.0646 & 0.9125 & 0.0747 & 0.9192 & 0.0794 & 0.9250 & 0.0823 \\
0.6 & 0.8853 & 0.0490 & 0.9027 & 0.0559 & 0.9087 & 0.0629 & 0.9148 & 0.0655 \\
0.7 & 0.8624 & 0.0408 & 0.8824 & 0.0455 & 0.8925 & 0.0492 & 0.8998 & 0.0507 \\
0.8 & 0.8253 & 0.0351 & 0.8499 & 0.0394 & 0.8618 & 0.0408 & 0.8711 & 0.0435 \\
0.9 & 0.7804 & 0.0287 & 0.8076 & 0.0333 & 0.8230 & 0.0359 & 0.8314 & 0.0377 \\
1.0 & 0.7306 & 0.0235 & 0.7610 & 0.0272 & 0.7740 & 0.0287 & 0.7822 & 0.0310 \\
\bottomrule
\end{tabular}
\caption{ExcluIR beta checkpoint results for Qwen/Qwen3-Embedding-4B with the target-minus-trap score. Columns report recall (R) and trap violation (V) at $k \in \{3,5,7,9\}$; rows use 0.10 beta checkpoints. The score is $S(d)=s(d,q_{\mathrm{target}})-\beta s(d,q_{\mathrm{trap}})$.}
\label{tab:appendix-excluir-target-qwen3-4b-beta-grid}
\endgroup
\end{table*}

\begin{table*}[t]
\centering
\begingroup
\small
\setlength{\tabcolsep}{4.2pt}
\renewcommand{\arraystretch}{1.05}
\begin{tabular}{@{}r*{8}{c}@{}}
\toprule
$\beta$ & \multicolumn{2}{c}{$k=3$} & \multicolumn{2}{c}{$k=5$} & \multicolumn{2}{c}{$k=7$} & \multicolumn{2}{c}{$k=9$} \\
\cmidrule(lr){2-3} \cmidrule(lr){4-5} \cmidrule(lr){6-7} \cmidrule(lr){8-9}
 & R & V & R & V & R & V & R & V \\
\midrule
0.0 & 0.9567 & 0.9814 & 0.9721 & 0.9938 & 0.9783 & 0.9969 & 0.9845 & 0.9969 \\
0.1 & 0.9598 & 0.9752 & 0.9721 & 0.9876 & 0.9752 & 0.9876 & 0.9876 & 0.9938 \\
0.2 & 0.9567 & 0.9659 & 0.9659 & 0.9783 & 0.9690 & 0.9845 & 0.9814 & 0.9876 \\
0.3 & 0.9474 & 0.9474 & 0.9628 & 0.9659 & 0.9659 & 0.9752 & 0.9690 & 0.9752 \\
0.4 & 0.9474 & 0.9226 & 0.9628 & 0.9536 & 0.9659 & 0.9628 & 0.9690 & 0.9659 \\
0.5 & 0.9412 & 0.8638 & 0.9567 & 0.9102 & 0.9598 & 0.9319 & 0.9628 & 0.9443 \\
0.6 & 0.9288 & 0.7833 & 0.9536 & 0.8359 & 0.9567 & 0.8638 & 0.9598 & 0.8762 \\
0.7 & 0.9040 & 0.6656 & 0.9195 & 0.7399 & 0.9288 & 0.7833 & 0.9412 & 0.7895 \\
0.8 & 0.8669 & 0.4211 & 0.8885 & 0.5046 & 0.9009 & 0.5418 & 0.9133 & 0.6037 \\
0.9 & 0.7740 & 0.1765 & 0.8050 & 0.2260 & 0.8235 & 0.2570 & 0.8452 & 0.2724 \\
1.0 & 0.6192 & 0.0836 & 0.6563 & 0.1176 & 0.6935 & 0.1269 & 0.7121 & 0.1269 \\
\bottomrule
\end{tabular}
\caption{BoolQuestions-not beta checkpoint results for text-embedding-3-small with the baseline-minus-trap score. Columns report recall (R) and trap violation (V) at $k \in \{3,5,7,9\}$; rows use 0.10 beta checkpoints. The score is $S(d)=s(d,q)-\beta s(d,q_{\mathrm{trap}})$.}
\label{tab:appendix-boolquestions-baseline-text-embedding-3-small-beta-grid}
\endgroup
\end{table*}

\begin{table*}[t]
\centering
\begingroup
\small
\setlength{\tabcolsep}{4.2pt}
\renewcommand{\arraystretch}{1.05}
\begin{tabular}{@{}r*{8}{c}@{}}
\toprule
$\beta$ & \multicolumn{2}{c}{$k=3$} & \multicolumn{2}{c}{$k=5$} & \multicolumn{2}{c}{$k=7$} & \multicolumn{2}{c}{$k=9$} \\
\cmidrule(lr){2-3} \cmidrule(lr){4-5} \cmidrule(lr){6-7} \cmidrule(lr){8-9}
 & R & V & R & V & R & V & R & V \\
\midrule
0.0 & 0.9536 & 0.9907 & 0.9752 & 0.9969 & 0.9783 & 0.9969 & 0.9876 & 0.9969 \\
0.1 & 0.9567 & 0.9845 & 0.9752 & 0.9969 & 0.9783 & 0.9969 & 0.9876 & 0.9969 \\
0.2 & 0.9536 & 0.9752 & 0.9783 & 0.9907 & 0.9876 & 0.9969 & 0.9876 & 0.9969 \\
0.3 & 0.9536 & 0.9381 & 0.9721 & 0.9690 & 0.9814 & 0.9690 & 0.9845 & 0.9721 \\
0.4 & 0.9505 & 0.9164 & 0.9752 & 0.9505 & 0.9814 & 0.9567 & 0.9814 & 0.9567 \\
0.5 & 0.9474 & 0.8885 & 0.9659 & 0.9164 & 0.9721 & 0.9350 & 0.9752 & 0.9443 \\
0.6 & 0.9257 & 0.7895 & 0.9567 & 0.8700 & 0.9690 & 0.8885 & 0.9690 & 0.8916 \\
0.7 & 0.9195 & 0.6780 & 0.9319 & 0.7492 & 0.9443 & 0.7802 & 0.9567 & 0.8019 \\
0.8 & 0.8824 & 0.4768 & 0.9009 & 0.5697 & 0.9257 & 0.6254 & 0.9350 & 0.6502 \\
0.9 & 0.7771 & 0.1950 & 0.8266 & 0.2508 & 0.8452 & 0.3189 & 0.8514 & 0.3375 \\
1.0 & 0.6440 & 0.0836 & 0.6904 & 0.0960 & 0.7121 & 0.1053 & 0.7368 & 0.1084 \\
\bottomrule
\end{tabular}
\caption{BoolQuestions-not beta checkpoint results for text-embedding-3-large with the baseline-minus-trap score. Columns report recall (R) and trap violation (V) at $k \in \{3,5,7,9\}$; rows use 0.10 beta checkpoints. The score is $S(d)=s(d,q)-\beta s(d,q_{\mathrm{trap}})$.}
\label{tab:appendix-boolquestions-baseline-text-embedding-3-large-beta-grid}
\endgroup
\end{table*}

\begin{table*}[t]
\centering
\begingroup
\small
\setlength{\tabcolsep}{4.2pt}
\renewcommand{\arraystretch}{1.05}
\begin{tabular}{@{}r*{8}{c}@{}}
\toprule
$\beta$ & \multicolumn{2}{c}{$k=3$} & \multicolumn{2}{c}{$k=5$} & \multicolumn{2}{c}{$k=7$} & \multicolumn{2}{c}{$k=9$} \\
\cmidrule(lr){2-3} \cmidrule(lr){4-5} \cmidrule(lr){6-7} \cmidrule(lr){8-9}
 & R & V & R & V & R & V & R & V \\
\midrule
0.0 & 0.9505 & 0.9845 & 0.9814 & 1.0000 & 0.9814 & 1.0000 & 0.9876 & 1.0000 \\
0.1 & 0.9443 & 0.9814 & 0.9752 & 0.9876 & 0.9845 & 0.9969 & 0.9845 & 1.0000 \\
0.2 & 0.9474 & 0.9659 & 0.9721 & 0.9845 & 0.9783 & 0.9845 & 0.9876 & 0.9907 \\
0.3 & 0.9474 & 0.9443 & 0.9598 & 0.9659 & 0.9721 & 0.9752 & 0.9814 & 0.9752 \\
0.4 & 0.9350 & 0.9102 & 0.9567 & 0.9443 & 0.9628 & 0.9567 & 0.9752 & 0.9628 \\
0.5 & 0.9257 & 0.8514 & 0.9443 & 0.8947 & 0.9598 & 0.9102 & 0.9659 & 0.9226 \\
0.6 & 0.9133 & 0.7864 & 0.9288 & 0.8452 & 0.9443 & 0.8607 & 0.9536 & 0.8793 \\
0.7 & 0.8947 & 0.6563 & 0.9164 & 0.7245 & 0.9288 & 0.7647 & 0.9350 & 0.7864 \\
0.8 & 0.8545 & 0.4272 & 0.8762 & 0.5139 & 0.8916 & 0.5697 & 0.9040 & 0.6037 \\
0.9 & 0.7833 & 0.1734 & 0.8173 & 0.2508 & 0.8266 & 0.2663 & 0.8390 & 0.3096 \\
1.0 & 0.6935 & 0.0898 & 0.7399 & 0.1084 & 0.7492 & 0.1269 & 0.7585 & 0.1331 \\
\bottomrule
\end{tabular}
\caption{BoolQuestions-not beta checkpoint results for Qwen/Qwen3-Embedding-0.6B with the baseline-minus-trap score. Columns report recall (R) and trap violation (V) at $k \in \{3,5,7,9\}$; rows use 0.10 beta checkpoints. The score is $S(d)=s(d,q)-\beta s(d,q_{\mathrm{trap}})$.}
\label{tab:appendix-boolquestions-baseline-qwen3-0p6b-beta-grid}
\endgroup
\end{table*}

\begin{table*}[t]
\centering
\begingroup
\small
\setlength{\tabcolsep}{4.2pt}
\renewcommand{\arraystretch}{1.05}
\begin{tabular}{@{}r*{8}{c}@{}}
\toprule
$\beta$ & \multicolumn{2}{c}{$k=3$} & \multicolumn{2}{c}{$k=5$} & \multicolumn{2}{c}{$k=7$} & \multicolumn{2}{c}{$k=9$} \\
\cmidrule(lr){2-3} \cmidrule(lr){4-5} \cmidrule(lr){6-7} \cmidrule(lr){8-9}
 & R & V & R & V & R & V & R & V \\
\midrule
0.0 & 0.9536 & 0.9783 & 0.9721 & 0.9876 & 0.9783 & 0.9907 & 0.9814 & 0.9938 \\
0.1 & 0.9536 & 0.9721 & 0.9721 & 0.9814 & 0.9752 & 0.9845 & 0.9814 & 0.9845 \\
0.2 & 0.9474 & 0.9505 & 0.9721 & 0.9814 & 0.9752 & 0.9814 & 0.9814 & 0.9814 \\
0.3 & 0.9505 & 0.9319 & 0.9659 & 0.9536 & 0.9690 & 0.9628 & 0.9752 & 0.9659 \\
0.4 & 0.9474 & 0.8885 & 0.9659 & 0.9257 & 0.9690 & 0.9412 & 0.9752 & 0.9412 \\
0.5 & 0.9474 & 0.8235 & 0.9598 & 0.8700 & 0.9628 & 0.8793 & 0.9659 & 0.8978 \\
0.6 & 0.9443 & 0.7771 & 0.9505 & 0.8111 & 0.9598 & 0.8297 & 0.9659 & 0.8390 \\
0.7 & 0.9195 & 0.6687 & 0.9381 & 0.7430 & 0.9443 & 0.7678 & 0.9474 & 0.7802 \\
0.8 & 0.8824 & 0.4768 & 0.9040 & 0.5480 & 0.9257 & 0.5975 & 0.9319 & 0.6471 \\
0.9 & 0.8111 & 0.1950 & 0.8607 & 0.2539 & 0.8700 & 0.2941 & 0.8824 & 0.3406 \\
1.0 & 0.7430 & 0.0867 & 0.7585 & 0.0898 & 0.7709 & 0.1022 & 0.7833 & 0.1053 \\
\bottomrule
\end{tabular}
\caption{BoolQuestions-not beta checkpoint results for Qwen/Qwen3-Embedding-4B with the baseline-minus-trap score. Columns report recall (R) and trap violation (V) at $k \in \{3,5,7,9\}$; rows use 0.10 beta checkpoints. The score is $S(d)=s(d,q)-\beta s(d,q_{\mathrm{trap}})$.}
\label{tab:appendix-boolquestions-baseline-qwen3-4b-beta-grid}
\endgroup
\end{table*}

\begin{table*}[t]
\centering
\begingroup
\small
\setlength{\tabcolsep}{4.2pt}
\renewcommand{\arraystretch}{1.05}
\begin{tabular}{@{}r*{8}{c}@{}}
\toprule
$\beta$ & \multicolumn{2}{c}{$k=3$} & \multicolumn{2}{c}{$k=5$} & \multicolumn{2}{c}{$k=7$} & \multicolumn{2}{c}{$k=9$} \\
\cmidrule(lr){2-3} \cmidrule(lr){4-5} \cmidrule(lr){6-7} \cmidrule(lr){8-9}
 & R & V & R & V & R & V & R & V \\
\midrule
0.0 & 0.9412 & 0.8545 & 0.9690 & 0.8947 & 0.9752 & 0.9071 & 0.9814 & 0.9195 \\
0.1 & 0.9381 & 0.8173 & 0.9628 & 0.8638 & 0.9721 & 0.8731 & 0.9783 & 0.8854 \\
0.2 & 0.9350 & 0.7771 & 0.9598 & 0.8204 & 0.9721 & 0.8452 & 0.9752 & 0.8545 \\
0.3 & 0.9226 & 0.7337 & 0.9443 & 0.7771 & 0.9598 & 0.7988 & 0.9659 & 0.8080 \\
0.4 & 0.9164 & 0.6873 & 0.9381 & 0.7245 & 0.9443 & 0.7461 & 0.9536 & 0.7616 \\
0.5 & 0.9071 & 0.6006 & 0.9288 & 0.6533 & 0.9412 & 0.6780 & 0.9443 & 0.6935 \\
0.6 & 0.8947 & 0.4799 & 0.9164 & 0.5480 & 0.9288 & 0.5728 & 0.9288 & 0.6161 \\
0.7 & 0.8700 & 0.3653 & 0.8854 & 0.4211 & 0.9009 & 0.4458 & 0.9102 & 0.4768 \\
0.8 & 0.8328 & 0.1920 & 0.8607 & 0.2570 & 0.8793 & 0.2848 & 0.8854 & 0.3220 \\
0.9 & 0.7090 & 0.1084 & 0.7492 & 0.1331 & 0.7895 & 0.1455 & 0.8050 & 0.1672 \\
1.0 & 0.5944 & 0.0557 & 0.6440 & 0.0774 & 0.6780 & 0.0836 & 0.6966 & 0.0898 \\
\bottomrule
\end{tabular}
\caption{BoolQuestions-not beta checkpoint results for text-embedding-3-small with the target-minus-trap score. Columns report recall (R) and trap violation (V) at $k \in \{3,5,7,9\}$; rows use 0.10 beta checkpoints. The score is $S(d)=s(d,q_{\mathrm{target}})-\beta s(d,q_{\mathrm{trap}})$.}
\label{tab:appendix-boolquestions-target-text-embedding-3-small-beta-grid}
\endgroup
\end{table*}

\begin{table*}[t]
\centering
\begingroup
\small
\setlength{\tabcolsep}{4.2pt}
\renewcommand{\arraystretch}{1.05}
\begin{tabular}{@{}r*{8}{c}@{}}
\toprule
$\beta$ & \multicolumn{2}{c}{$k=3$} & \multicolumn{2}{c}{$k=5$} & \multicolumn{2}{c}{$k=7$} & \multicolumn{2}{c}{$k=9$} \\
\cmidrule(lr){2-3} \cmidrule(lr){4-5} \cmidrule(lr){6-7} \cmidrule(lr){8-9}
 & R & V & R & V & R & V & R & V \\
\midrule
0.0 & 0.9536 & 0.8731 & 0.9721 & 0.9071 & 0.9845 & 0.9164 & 0.9876 & 0.9319 \\
0.1 & 0.9536 & 0.8297 & 0.9690 & 0.8793 & 0.9845 & 0.8947 & 0.9845 & 0.9009 \\
0.2 & 0.9474 & 0.7957 & 0.9690 & 0.8390 & 0.9845 & 0.8545 & 0.9845 & 0.8669 \\
0.3 & 0.9474 & 0.7368 & 0.9659 & 0.7957 & 0.9783 & 0.8111 & 0.9845 & 0.8235 \\
0.4 & 0.9412 & 0.6749 & 0.9628 & 0.7245 & 0.9721 & 0.7678 & 0.9845 & 0.7833 \\
0.5 & 0.9226 & 0.6192 & 0.9536 & 0.6656 & 0.9628 & 0.6842 & 0.9690 & 0.7028 \\
0.6 & 0.9195 & 0.5356 & 0.9350 & 0.5913 & 0.9474 & 0.6254 & 0.9598 & 0.6378 \\
0.7 & 0.8793 & 0.4334 & 0.9071 & 0.4830 & 0.9288 & 0.5325 & 0.9381 & 0.5542 \\
0.8 & 0.8576 & 0.2632 & 0.8762 & 0.3220 & 0.8885 & 0.3715 & 0.8978 & 0.3901 \\
0.9 & 0.7678 & 0.1331 & 0.7926 & 0.1517 & 0.8173 & 0.1641 & 0.8390 & 0.1889 \\
1.0 & 0.6656 & 0.0805 & 0.6997 & 0.0836 & 0.7152 & 0.0836 & 0.7337 & 0.0836 \\
\bottomrule
\end{tabular}
\caption{BoolQuestions-not beta checkpoint results for text-embedding-3-large with the target-minus-trap score. Columns report recall (R) and trap violation (V) at $k \in \{3,5,7,9\}$; rows use 0.10 beta checkpoints. The score is $S(d)=s(d,q_{\mathrm{target}})-\beta s(d,q_{\mathrm{trap}})$.}
\label{tab:appendix-boolquestions-target-text-embedding-3-large-beta-grid}
\endgroup
\end{table*}

\begin{table*}[t]
\centering
\begingroup
\small
\setlength{\tabcolsep}{4.2pt}
\renewcommand{\arraystretch}{1.05}
\begin{tabular}{@{}r*{8}{c}@{}}
\toprule
$\beta$ & \multicolumn{2}{c}{$k=3$} & \multicolumn{2}{c}{$k=5$} & \multicolumn{2}{c}{$k=7$} & \multicolumn{2}{c}{$k=9$} \\
\cmidrule(lr){2-3} \cmidrule(lr){4-5} \cmidrule(lr){6-7} \cmidrule(lr){8-9}
 & R & V & R & V & R & V & R & V \\
\midrule
0.0 & 0.9598 & 0.8731 & 0.9721 & 0.9071 & 0.9752 & 0.9257 & 0.9783 & 0.9319 \\
0.1 & 0.9536 & 0.8142 & 0.9690 & 0.8576 & 0.9721 & 0.8824 & 0.9752 & 0.9071 \\
0.2 & 0.9474 & 0.7616 & 0.9659 & 0.8173 & 0.9721 & 0.8421 & 0.9721 & 0.8483 \\
0.3 & 0.9381 & 0.6997 & 0.9598 & 0.7647 & 0.9628 & 0.7833 & 0.9659 & 0.7957 \\
0.4 & 0.9319 & 0.6409 & 0.9443 & 0.6873 & 0.9536 & 0.7121 & 0.9598 & 0.7399 \\
0.5 & 0.9164 & 0.5666 & 0.9350 & 0.6161 & 0.9412 & 0.6409 & 0.9505 & 0.6594 \\
0.6 & 0.9102 & 0.4520 & 0.9195 & 0.5201 & 0.9195 & 0.5573 & 0.9381 & 0.5789 \\
0.7 & 0.8854 & 0.3375 & 0.8978 & 0.3746 & 0.9040 & 0.4087 & 0.9102 & 0.4396 \\
0.8 & 0.8359 & 0.1920 & 0.8576 & 0.2291 & 0.8607 & 0.2570 & 0.8700 & 0.2724 \\
0.9 & 0.7864 & 0.1115 & 0.8173 & 0.1238 & 0.8297 & 0.1393 & 0.8328 & 0.1548 \\
1.0 & 0.6935 & 0.0588 & 0.7368 & 0.0805 & 0.7461 & 0.0929 & 0.7709 & 0.0929 \\
\bottomrule
\end{tabular}
\caption{BoolQuestions-not beta checkpoint results for Qwen/Qwen3-Embedding-0.6B with the target-minus-trap score. Columns report recall (R) and trap violation (V) at $k \in \{3,5,7,9\}$; rows use 0.10 beta checkpoints. The score is $S(d)=s(d,q_{\mathrm{target}})-\beta s(d,q_{\mathrm{trap}})$.}
\label{tab:appendix-boolquestions-target-qwen3-0p6b-beta-grid}
\endgroup
\end{table*}

\begin{table*}[t]
\centering
\begingroup
\small
\setlength{\tabcolsep}{4.2pt}
\renewcommand{\arraystretch}{1.05}
\begin{tabular}{@{}r*{8}{c}@{}}
\toprule
$\beta$ & \multicolumn{2}{c}{$k=3$} & \multicolumn{2}{c}{$k=5$} & \multicolumn{2}{c}{$k=7$} & \multicolumn{2}{c}{$k=9$} \\
\cmidrule(lr){2-3} \cmidrule(lr){4-5} \cmidrule(lr){6-7} \cmidrule(lr){8-9}
 & R & V & R & V & R & V & R & V \\
\midrule
0.0 & 0.9567 & 0.8669 & 0.9783 & 0.9133 & 0.9845 & 0.9226 & 0.9845 & 0.9288 \\
0.1 & 0.9628 & 0.8142 & 0.9752 & 0.8700 & 0.9845 & 0.8885 & 0.9845 & 0.9009 \\
0.2 & 0.9598 & 0.7678 & 0.9659 & 0.8204 & 0.9752 & 0.8421 & 0.9845 & 0.8514 \\
0.3 & 0.9536 & 0.7276 & 0.9659 & 0.7709 & 0.9690 & 0.7802 & 0.9783 & 0.7988 \\
0.4 & 0.9350 & 0.6842 & 0.9628 & 0.7183 & 0.9690 & 0.7399 & 0.9690 & 0.7585 \\
0.5 & 0.9319 & 0.5913 & 0.9505 & 0.6440 & 0.9598 & 0.6811 & 0.9628 & 0.6966 \\
0.6 & 0.9257 & 0.4830 & 0.9350 & 0.5232 & 0.9412 & 0.5851 & 0.9505 & 0.6130 \\
0.7 & 0.8978 & 0.3375 & 0.9164 & 0.4025 & 0.9319 & 0.4458 & 0.9350 & 0.4613 \\
0.8 & 0.8576 & 0.1858 & 0.8793 & 0.2415 & 0.8916 & 0.2817 & 0.9071 & 0.2972 \\
0.9 & 0.8019 & 0.1053 & 0.8204 & 0.1207 & 0.8421 & 0.1393 & 0.8514 & 0.1579 \\
1.0 & 0.7337 & 0.0588 & 0.7585 & 0.0681 & 0.7802 & 0.0712 & 0.7988 & 0.0774 \\
\bottomrule
\end{tabular}
\caption{BoolQuestions-not beta checkpoint results for Qwen/Qwen3-Embedding-4B with the target-minus-trap score. Columns report recall (R) and trap violation (V) at $k \in \{3,5,7,9\}$; rows use 0.10 beta checkpoints. The score is $S(d)=s(d,q_{\mathrm{target}})-\beta s(d,q_{\mathrm{trap}})$.}
\label{tab:appendix-boolquestions-target-qwen3-4b-beta-grid}
\endgroup
\end{table*}

\end{document}